\documentclass[aps,prb,superscriptaddress,showpacs,floatfix,twocolumn]{revtex4-1}

\usepackage{times}
\usepackage{graphicx,graphics,color}% Include figure files
\usepackage{amssymb}
\usepackage{bm}
\usepackage{amsmath}
\usepackage{gensymb}
\usepackage{caption3}
\usepackage{subfigure}
\newcommand{\Px}	{BaFe$_{2}$(As$_{1-x}$P$_{x}$)$_{2}$\,}
\newcommand{\BFA}	{BaFe$_{2}$As$_{2}$\,}
\newcommand{\BFP}	{BaFe$_{2}$P$_{2}$\,}
\newcommand{\MAG}	{$\langle |M|\rangle$\,}
\newcommand{\DEL}	{$\langle |\Delta_s|\rangle$\,}

\begin{document}

\title{Phase diagram of the isovalent phosphorous-substituted 122-type iron pnictides}

\author{YuanYuan Zhao}
\affiliation{Department of Physics and Texas Center for Superconductivity, University of Houston, Houston, Texas 77204, USA}

\author{Yuan-Yen Tai}
\affiliation{Theoretical Division, Los Alamos National Laboratory, Los Alamos, New Mexico 87545, USA}

\author{C. S. Ting}
\affiliation{Department of Physics and Texas Center for Superconductivity, University of Houston, Houston, Texas 77204, USA}

\date{\today}

\begin{abstract}

Recent experiments demonstrated that isovalent doping system gives the similar phase diagram as the heterovalent doped cases.
For example, with the phosphorous (P)-doping, the magnetic order in \Px compound is first suppressed, then the superconductivity dome emerges to an extended doping region but eventually it disappears at large $x$.
With the help of a minimal two-orbital model for both \BFA and \BFP, together with the self-consistent lattice Bogoliubov-de Gennes (BdG) equation, we calculate the phase diagram against the P content $x$ in which the doped isovalent P-atoms are treated as impurities.
We show that our numerical results can qualitatively compare with the experimental measurements.

\end{abstract}

\maketitle

\section{Introduction}
Recent intensive studies on the 122 family of iron-pnictide XFe$_{2}$As$_{2}$ (where X = Ba, Sr or Ca) compounds found that the superconductivity (SC) can be induced by different means~\cite{Rotter2008L, Sefat2008, Alireza2009, Sharma2010, Jiang2009}. In all these cases, the resulting phase diagrams are quite similar. Starting from a co-linear spin-density wave (SDW) metal, a SC dome emerges with doping in the parent compound, while the SDW gets suppressed. In the electron-doped and hole-doped materials, the emergence of SC is due to the imbalance between the electron and hole carrier densities in the system. However, isovalent substitution of As with phosphorous (P) in \BFA shows similar phase diagram~\cite{Jiang2009, Kasahara2010, Nakai2010, Bohmer2012} without introducing additional net charge carriers. This raises the question about the underlying mechanism for the phase transition from the SDW to SC, and eventually to normal metal as the doping parameter $x$ in the system increases, which is the issue we wish to address in the present work.

Since P anion is smaller than As anion, it has been suggested that superconductivity in this isovalent system is induced by a chemical pressure~\cite{Jiang2009, Kasahara2010, Klintberg2010}, or it is correlated with the distinct role of lattice parameters, \textit{e.g.}, the As-Fe-As bond angle~\cite{CHLee2008, Liwei2012}, anion-height~\cite{Mizuguchi2010}, and bond length of Fe-As and Fe-P~\cite{Zinth2012}.
Besides, it has also been suggested that the uniaxial pressure~\cite{Bohmer2012}, similar to the electron-doped Ba(Fe$_{1-x}$Co$_{x}$)$_{2}$As$_{2}$ system~\cite{Drotziger2010, Meingast2012}, and charge inhomogeneity~\cite{Feng2012} plays an important role in the emergence of superconductivity.
On the other hand, Rotter and co-authors~\cite{Rotter2010} compared structure data determined by high-resolution x-ray powder and single-crystal diffraction, with theoretical models obtained by DFT calculations, and emphasized that even subtle details of the crystal structures are crucial to magnetism and superconductivity.

In this work, we want to answer the outstanding question that ``What is the essential ingredient to cause the phase diagram of \Px?\,".
Before we dive into the detail of our minimal effective Hamiltonian, we need to address several points for how we choose the tight-binding model.
First, in order to understand the electronic structure of the compound \Px, the scatterings of charge carriers by the randomly doped P- atoms or sites should be carefully addressed.
Since the impurity (or P-atom) concentration are taking from $0$ to $100\%$ in the calculation, it is essential to construct a large real-space lattice to capture reasonable statistical ensemble.
Historically, several microscopic multi-orbital models have been developed~\cite{Raghu2008, Kuroki2009, Laad2009, Zhang2009, Wang2010, Thomale2011, Su2012, HuS4, YuanYenepl} and we adopt a minimal two-dimensional (2D) and two-orbital model from Tai and his co-authors~\cite{YuanYenepl}.
This model has been tested for capturing several important features of the 122 \BFA compounds, such as the paramagnetic band structure and Fermi surface topology according to the d$_{xz}$(d$_{yz}$) orbital ordering, the entire hole- and electron-doped phase diagram~\cite{YuanYenepl}, the evolution of impurity quasiparticle states~\cite{LPan1} and the Fermi surface evolution for the co-linear SDW phase~\cite{LPan2}. All these features are in agreement with experiments.
More importantly, our effective two-orbital model helps reducing the degree of freedom from the complexity associated with the multi-orbital models and enables us to construct a large lattice Hamiltonian to perform a cost-effective calculation in the presence of disorders and to calculate the phase diagram as a function of the P-doping.  It is almost impossible to apply the three-orbital (d$_{xz}$, d$_{yz}$ and d$_{xy}$), the five d-orbital and the eight-orbital (Fe-d+As-p) models to perform the same type of calculation~\cite{footnote01}. Afterall, the essential features of the magnetic and SC orders at arbitrary doping could be accounted for by d$_{xz}$ and d$_{yz}$ orbitals. The other orbitals only make minor and quantitative modifications.

\section{Model construction}
We begin with \BFA, a prototype parent compound of superconductors, showing a co-linear SDW antiferromagnetic ground state~\cite{Rotter2008B}.
The Fe atoms in \BFA form a square lattice, while the As atoms sit in the center of each square plaquette of the Fe lattice and are displaced alternatively above and below the Fe-Fe plane, which leads to two sublattices of Fe atoms denoted by sublattice $A$ and $B$.
In Ref.~\onlinecite{YuanYenepl}, the authors proposed a minimal 2-orbital model with these two Fe atoms per unit-cell through considering the orbital ordering physics of Fe-$3d_{xz}$ and Fe-$3d_{yz}$ orbitals, and later on it has been proven that one could also use a gauge transform to represent this model within 1-Fe atom per unit-cell ~\cite{HuaChenprb}.

Here we start with the kinetic term of the lattice Hamiltonian for the mixed compound, \Px, written as, $H_0 = H_{\tilde t} + H_{P_{inter}}$.
Where $H_{\tilde t}$ is the hopping term in real space and the {\it tilde} symbol of $\tilde t$ is presenting either for pure \BFA, pure \BFP, or the different ones caused by the As-P mixture,
\begin{equation}\label{HamHop}
    H_{\tilde t} = \sum_{i, j, \alpha, \beta, \sigma} {\tilde t}_{i\, j}^{\alpha \, \beta} c_{i\, \alpha \, \sigma}^{\dag} c_{j\, \beta \, \sigma} - \sum_{i, \alpha, \sigma} \mu\, c_{i\, \alpha \, \sigma}^{\dag} c_{i\, \alpha \, \sigma}
\end{equation}
where $c_{i\, \alpha \, \sigma}^{\dag}$ and $c_{i\, \alpha \, \sigma}$ are respectively the creation and annihilation operators for an electron with spin $\sigma$ in the orbitals $\alpha = 1$ or $2$ on the $i$-th lattice site, $\mu$ is the chemical potential which is adjusted to give a fixed filling factor, and $\tilde t_{i\, j}^{\alpha \, \beta}$ are the hopping integrals.
We choose the nonvanishing hopping elements as~\cite{YuanYenepl,HuaChenprb}: $\tilde t^{\alpha\bar\alpha}_{\pm\hat x} = \tilde t^{\alpha\bar\alpha}_{\pm\hat y}=\tilde t_1$, $\tilde t^{11}_{\pm(\hat x+\hat y)} = \tilde t^{22}_{\pm(\hat x-\hat y)}=\tilde t_2$, $\tilde t^{11}_{\pm(\hat x-\hat y)} = \tilde t^{22}_{\pm(\hat x+\hat y)}=\tilde t_3$, $\tilde t^{\alpha\bar\alpha}_{\pm(\hat x\pm\hat y)}=\tilde t_4$, $\tilde t^{\alpha\alpha}_{\pm\hat x} =\tilde  t^{\alpha\alpha}_{\pm\hat y}=\tilde t_5$, $\tilde t^{\alpha\alpha}_{\pm2\hat x}=\tilde t^{\alpha\alpha}_{\pm2\hat y}=\tilde t_6$.
The second term in $H_0$ is on-site interorbital scattering of the four Fe sublattices around the substituted-$P$ atom,
$H_{P_{inter}} = \sum_{i, \alpha \neq \beta, \sigma} V_{inter} c_{i\, \alpha \, \sigma}^{\dag} c_{i\, \beta \, \sigma}$
here $V_{inter} = 0.031$, describes a very weak on-site interorbital scattering, and we assume it is caused by the existence of substituted-$P$ atoms.
It was found the on-site interorbital scattering will suppress, and even completely destroy the superconductivity in the system~\cite{Ciechan2009}.
Here, we choose the six hopping integrals, $\tilde t$, to be:
\begin{equation}
\begin{aligned}
\label{hoppings}
\tilde t_{1-6}	    &=t_{1-6} = ({\bf 0.09,\, 0.08,\, 1.35,\, -0.12,\, -1.00,\, 0.25}),\\
				    &\mbox{for pure \BFA system (Appendix in Ref.~\cite{HuaChenprb})};\\
\tilde t_{1-6}	    &=t^{\prime}_{1-6} = ({\bf 0.25,\, 0.35,\, 1.65,\, -0.22,\, -1.45,\, 0.25}),\\
				    &\mbox{for pure \BFP system};\\
\tilde t_{1,\,5}    &=\tilde{a}\, t_{1,5}+\tilde{b}\, t^{\prime}_{1,5},\;\; \mbox{(two sets with $\tilde{a}+\tilde{b}=1$; $\tilde{a},\tilde{b}>0$),}\\
				    & \mbox{for the NN-$t$ on the As-P boundary};\\
\tilde t_{2,\,3,\,4}&=a\, t_{2,\,3,\,4}+b\, t^{\prime}_{2,\,3,\,4},\;\; \mbox{($a+b=1$; $a,b>0$),}\\
				    & \mbox{for the NNN-$t$ around substituted-P.}
\end{aligned}
\end{equation}
where NN represents the nearest-neighbor, and NNN represents the next-nearest-neighbor.
Note that, with substituted-P atoms in the system, which is a mixing of As and P atoms, we propose a mixed hopping integrals varying caused by the As-P mixture as shown in Eq.~\ref{hoppings}.
The $a$ and $b$ values may vary according to different situations, the technical details of the mixing in hopping integrals are given in the Appendix A.%~\ref{suphop}.

Now, we are ready to write down the full Hamiltonian with the modified hopping term $H_0$ for \Px,
\begin{equation}
\label{FullHam}
H = H_0 + H_{int} + H_{\Delta}.
\end{equation}

We follow the same formalism as in Ref.~\onlinecite{HuaChenprb} for the interaction terms, $H_{int}$ and $H_{\Delta}$. Where $H_{int}$ contains an on-site Hubbard, $U$, and Hund's coupling, $J_{H}$, which is responsible for the co-linear SDW order.
\begin{equation}
\begin{aligned}
\label{HamInt}
         H_{int}&= U \sum_{i, \alpha, \sigma} \langle \hat{n}_{i\, \alpha \, \bar{\sigma}} \rangle \, \hat{n}_{i\, \alpha \, \sigma}
           + U^{\prime} \sum_{i, \alpha \neq \beta, \sigma \neq \bar{\sigma}} \langle \hat{n}_{i\, \alpha \, \bar{\sigma}} \rangle \, \hat{n}_{i\, \beta \, \sigma} \\
            & + (U^{\prime} - J_{H}) \sum_{i, \alpha \neq \beta, \sigma} \langle \hat{n}_{i\, \alpha \, \sigma} \rangle \, \hat{n}_{i\, \beta \, \sigma}
\end{aligned}
\end{equation}
where $\hat{n}_{i\, \alpha \, \sigma} = c_{i\, \alpha \, \sigma}^{\dag} \, c_{i\, \alpha \, \sigma}$. The orbital rotation symmetry imposes the constraint $U = U^{\prime} + 2\,J_{H}$.

Here our pairing interaction is represented by $H_{\Delta}=-\sum_{ij\sigma}V_{ij} n_{i\alpha\downarrow} n_{j\alpha\uparrow}$ with $V_{ij}>0$.
The mean-field decoupling of  $H_{\Delta}$ can be written in the following form,
\begin{equation}
\label{HamInt}
H_{\Delta} = \sum_{i, j, \alpha} V_{ij} \left( \langle c_{i\, \alpha \, \downarrow} c_{j\, \alpha \, \uparrow}\rangle  c_{i\, \alpha \, \downarrow}^{\dag} c_{j\, \alpha \, \uparrow}^{\dag} + H.c. \right).
\end{equation}

$V_{ij} \langle c_{i\, \alpha \, \downarrow} c_{j\, \alpha \,\uparrow}\rangle =\Delta_{i\,j}^{\alpha}$ is the SC bond pairing order parameter between site $i$ and site $j$. In principle, such an `{\it attractive}' interaction ($-V_{ij}$) between electrons in real space could be generated via the on-site Hubbard-$U$ interaction according to the spin-fluctuation theory in real-space, $V_{ij}\sim \chi_{ij}$, as described in Ref.~\onlinecite{ATRomer}, which makes a consistent picture as one considers the case of spin-fluctuation theorem in k-space~\cite{Graser2009}.
In this paper, we want to emphasis on the effect of the mixed hopping, $\tilde t$, and we do not address the full real-space RPA calculation of $V_{ij}$~\cite{ATRomer}. The nearest neighboring pairing interaction $V_{NN}$ is known to give rise to the $d$-wave pairing, while the next-nearest neighboring $V_{NNN}$ would be responsible for the s$_\pm$--pairing symmetry that clearly gaining experimental supports for the electron and hole-doped \BFA compounds. We believe that the d-wave component of the SC order parameter must be completely suppressed in the compound.  In this paper, we assume that SC of \Px has the s -pairing symmetry in the Fe-plane. Thus we only need to consider $V_{NNN}$ for the pairing interaction. The interplay between $V_{NN}$ and $V_{NNN}$ on the pairing symmetry of the electron-doped \BFA has also been recently studied by our group~\cite{BLi}. It was demonstrated that the $d$-wave component of the SC order parameter could easily be suppressed by the s$_\pm$-wave pairing component, as well as by the disordered scatterings due to the randomly distributed dopants in the Fe-planes.  Therefore, we only take the NNN pairing interaction $V_{NNN}$ into account and treated it effectively as a constant, and this interaction would generate a SC with the s$_\pm$-wave pairing symmetry. In our model calculation, we did not find that the charge-density wave state is stable~\cite{footnote03}.

We write down the matrix form of Eq.~\ref{FullHam} with basis $\psi_{i\alpha}=(c_{i\alpha\uparrow},c^\dagger_{i\alpha\downarrow})^{Transpose}$, $H=\sum_{ij\alpha\beta}\psi^\dagger_{i\alpha}\, H_{BdG} \,\psi_{j\beta}$,  and calculate the eigenvalue and eigenvector of $H_{BdG}$:
\begin{equation}\label{BdG}
    \sum_{j\,\beta}
    \left(\begin{array}{cc}
        H_{i\, j\, \uparrow}^{\alpha\, \beta} & \Delta_{i\,j}^{\beta}\\
        \\
        \Delta_{i\,j}^{\beta*}& - H_{i\, j\, \downarrow}^{\alpha \, \beta}
    \end{array} \right)
    \left(\begin{array}{c}
        u_{j\, \beta}^{n} \\
        \\
        v_{j\, \beta}^{n}
    \end{array} \right)
    = E_{n}
    \left(\begin{array}{c}
        u_{i\, \alpha}^{n} \\
        \\
        v_{i\, \alpha}^{n}
    \end{array} \right)
\end{equation}
where, $H_{i\, j\, \sigma}^{\alpha\, \beta} \equiv [ H_{0}+H_{int}]_{ij\sigma}^{\alpha\beta}$, is the matrix-element for the single-particle Hamiltonian.
We solve the mean-field order parameters
$\langle \hat{n}_{i\, \alpha \, \uparrow} \rangle = \sum_{n} \left| u_{i\, \alpha}^{n} \right|^{2} f(E_{n})$,
$\langle \hat{n}_{i\, \alpha \, \downarrow} \rangle = \sum_{n} \left| v_{i\, \alpha}^{n} \right|^{2} [1 - f(E_{n})]$ and
\begin{equation}
\Delta_{i\,j}^{\alpha} = \frac{V_{ij}}{4}\sum_{n}( u_{i\, \alpha}^{n} v_{j\, \alpha}^{n*}
						+ u_{j\, \alpha}^{n} v_{i\, \alpha}^{n*})\,\mbox{tanh}\big(\frac{E_n}{2k_B T}\big),
\end{equation}
self-consistently with Eq.~\ref{BdG}, where $f(E_{n})$ is the Fermi-Dirac distribution function.
As described above, we only consider the NNN intra-orbital pairing with pairing potential $V_{ij} = V_{\text{NNN}} = V$ to serve the $s\pm$ pairing symmetry. To facilitate the discussion of physical observables and generating of the phase diagram, we define respectively the staggered lattice magnetization and the $s$-wave projection of the SC order parameter at each site $i$, as:
$m_{i} = \frac{1}{4}\sum_{\alpha} (\langle \hat{n}_{i \alpha \uparrow} \rangle-\langle \hat{n}_{i \alpha \downarrow} \rangle)$,
$\Delta_{i}= \frac{1}{8}\sum_{\delta, \alpha} \Delta_{i\,i+\delta}^{\alpha}$, where $\delta\in\{\pm\hat x\pm\hat y\}$.
The neighbors of site $i$ are reached by $\delta$. Besides, we also calculate the averaged values $\langle |M|\rangle=\frac{1}{N}\sum_i |m_i|$ and $\langle |\Delta_s|\rangle=\frac{1}{N}\sum_i |\Delta_i|$, to investigate the phase diagram. $N$ is the number of Fe sites in the real-space lattice.

Throughout this paper, we choose fixed interaction parameters $( U, J_{H}, V ) = (3.2,\, 0.6,\, 1.05)$, no matter whether there are substituted-P atoms or not. In principal, $U$, $J_H$ and $V$ should be changed somewhat due to the substitution of P. If we do so, our result could be better fitted to the experiment. However, in the present study, we would like to focus our attention only to the hopping effect on the phase diagram without changing the interaction terms.

\section{Band structure and static spin susceptibility}
\begin{figure}[htbp]
\centering
\!\!\!\!
\subfigure[]{\includegraphics[width=0.4\linewidth, clip=true]{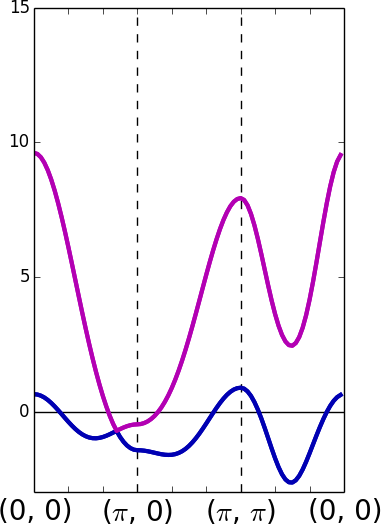}\label{bandAs}}
\qquad\quad
\subfigure[]{\includegraphics[width=0.4\linewidth, clip=true]{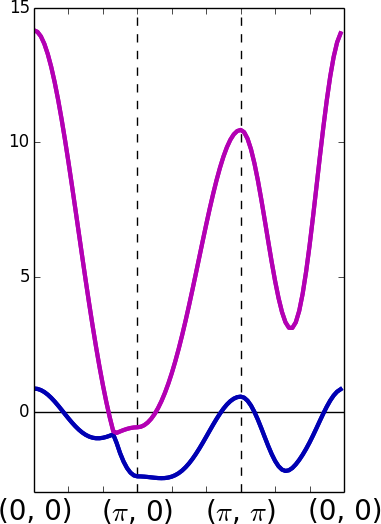}\label{bandP}}
\\
\subfigure[]{\includegraphics[width=0.42\linewidth, clip=true]{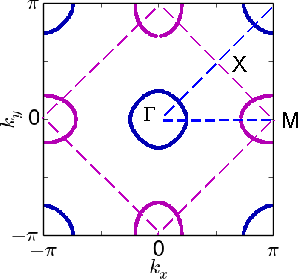}\label{fermiAs}}
\qquad\quad
\subfigure[]{\includegraphics[width=0.42\linewidth, clip=true]{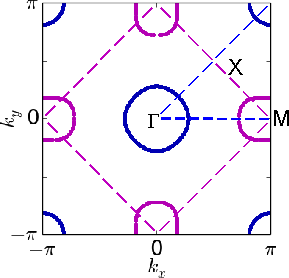}\label{fermiP}}
\caption{(Color online) The model calculated band structure of two-orbital model on the unfolded (1-Fe atom per unit cell) Brillouin Zone (BZ) for (a) \BFA ($x=0$), (b) \BFP ($x=1$); and Fermi surface on the unfolded BZ for (c) \BFA ($x=0$), (d) \BFP ($x=1$). Here the fermi energy is shifted to zero for $1/2$ filling.} \label{bandstr}
\end{figure}

In this section, we systematically study the band structure and Fermi surface for pure \BFA and \BFP.
We will also compare these features through experiments and LDA calculations from the literature.
Due to the periodicity, the real-space kinetic term without spin indices, $\sum \tilde t_{i\, j}^{\alpha \, \beta} c_{i\, \alpha}^{\dag} c_{j\, \beta}+ \sum_{i,\alpha}V_{inter} c^\dagger_{i\alpha} c_{i\bar\alpha}$, can be easily Fourier transformed to the {\bf k}-space, $H_{\tilde t} =\frac{1}{N} \sum_{\bf k} \phi^{\dag}_{\bf k} M_{\bf k} \phi_{\bf k}$, with the 1-Fe per unit cell basis.
Here $\phi_{\bf k}=(c_{1,\bf k}, c_{2,\bf k})^T$ and
\begin{equation}\label{TBHt}
    M_{\bf k} = \left( \begin{array}{cc}
        \epsilon_s+\zeta_{+} - \mu & \epsilon_{o}+V_{inter} \\
        \epsilon_{o}+V_{inter} & \epsilon_s+\zeta_{-} - \mu
    \end{array} \right)
\end{equation}
where,
\begin{eqnarray*}
    \epsilon_{s} &=& 2\, \tilde t_{1} \left( \cos{k_{x}} + \cos{k_{y}} \right) + 2\, \tilde t_{6} \left( \cos{2 k_{x}} + \cos{2 k_{y}} \right), \\
    \zeta_{+} &=& 2 \left( \tilde t_{2} + \tilde t_{3} \right) \cos{k_{x}} \cos{k_{y}} + 2 \left( \tilde t_{2} - \tilde t_{3} \right) \sin{k_{x}} \sin{k_{y}}, \\
    \zeta_{-} &=& 2 \left( \tilde t_{2} + \tilde t_{3} \right) \cos{k_{x}} \cos{k_{y}} - 2 \left( \tilde t_{2} - \tilde t_{3} \right) \sin{k_{x}} \sin{k_{y}}, \\
    \epsilon_{o} &=& 2\, \tilde t_{5} \left( \cos{k_{x}} + \cos{k_{y}} \right) + 4\, \tilde t_{4} \cos{k_{x}} \cos{k_{y}}.
\end{eqnarray*}
We diagonalize Eq.~\ref{TBHt} to obtain the electronic structures of pure \BFA ($\tilde t=t$; $V_{inter}=0$) and pure \BFP ($\tilde t=t'$; $V_{inter}=0.031$) compounds with the parameters given in Eq.~\ref{hoppings}. The band structures and Fermi surfaces of these two systems are shown in Fig.~\ref{bandstr} for the 1-Fe atom per unit cell Brillouin zone (BZ). Through the literature study, we found that the Fermi surface for \BFA and \BFP share very similar signature, \textit{e.g.}, the de Haas-van Alphen experiments~\cite{Kasahara2010, Shishido2010, Analytis2010}, the angle-resolved photoemission spectroscopy (ARPES) experiments~\cite{Yoshida2011} and the LDA extracted 10-orbital model~\cite{KSuzuki2011}. Here, from our model parameter, it can be seen from Fig.~\ref{bandstr} that the band structures and Fermi surfaces of these two compounds are similar to each other, however, the hole pocket around $\Gamma$ point becomes more expanded in \BFP than that in \BFA, which is also in good agreement with experiments~\cite{Shishido2010}. Moreover, the band width of \BFP is wider than that of \BFA, which is consistent with recent ARPES experiment~\cite{Feng2012}. More detailed band structures and Fermi surfaces based on the folded BZ of 2-Fe per unit cell are shown in the Appendix B.

\begin{figure}[htbp]
\subfigure[]{\includegraphics[width=0.48\linewidth]{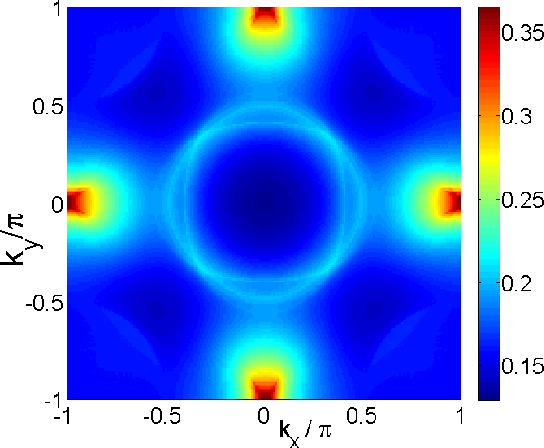}\label{susAs}}
\;
\subfigure[]{\includegraphics[width=0.48\linewidth]{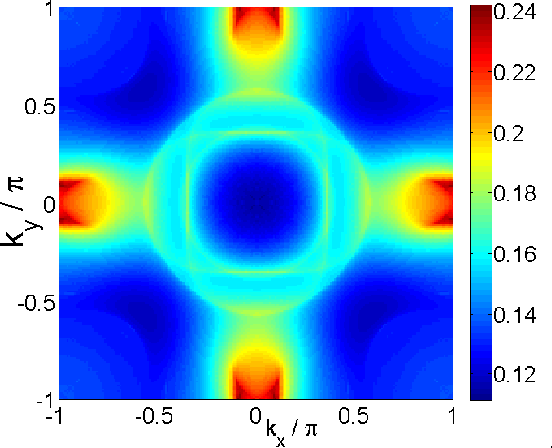}\label{susP}}
\\ \!\!\!\!
\subfigure[]{\includegraphics[width=0.45\linewidth]{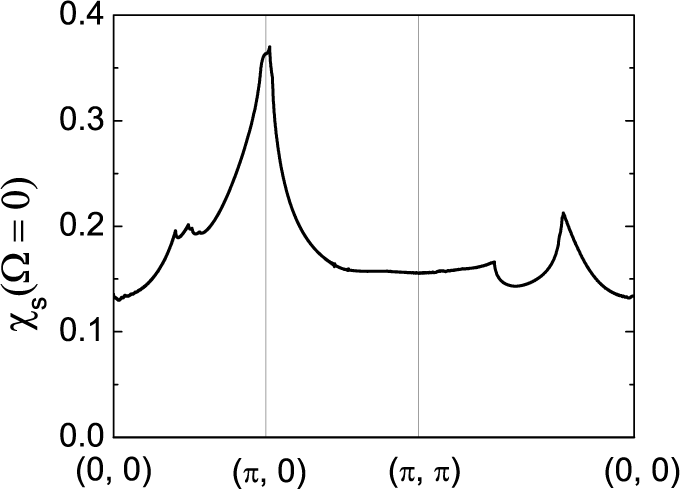}\label{susAs2D}}
\quad
\subfigure[]{\includegraphics[width=0.45\linewidth]{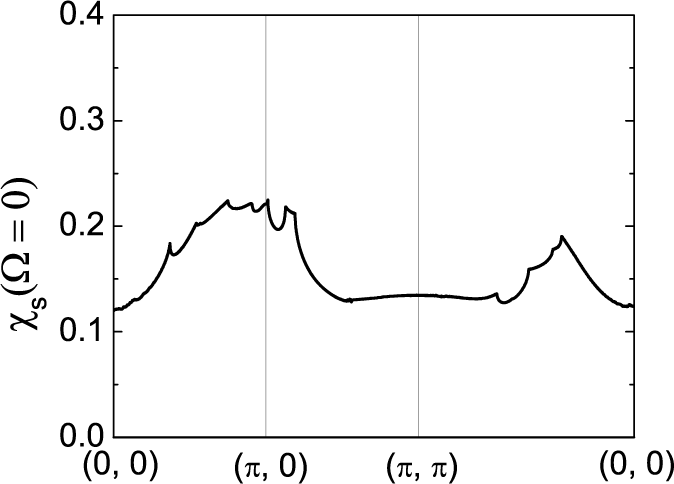}\label{susP2D}}
\caption{
(Color online) The bare spin susceptibility $\chi_{s}$ versus ${\bf q}$ with the tight-binding parameters as used in Fig.~\ref{bandstr} for (a, c) \BFA ($x=0$); (b, d) \BFP ($x=1$).
While (a) and (b) show the bare spin susceptibility in the first BZ, (c) and (d) show cuts of bare spin susceptibility along the main symmetry directions.
} \label{spinsus}
\end{figure}

Now, we study the static spin susceptibility bubble for the tight-binding model without electron-electron interaction. The static spin susceptibility can be obtained as,
\begin{equation}
\begin{aligned}
    \chi_{s}({\bf q}, i \Omega) =& -\frac{T}{2N}\sum_{{\bf k}, \omega_{n}} \textmd{Tr} \left[ G ({\bf k} + {\bf q}, \, i\omega_{n} + i \Omega) G ({\bf k}, \, i\omega_{n}) \right], \\
                                 =& -\frac{1}{2N} \sum_{{\bf k},\nu,\nu'} \frac{ |\langle {\bf k} + {\bf q},\nu\; \big| \; {\bf k},\nu'\rangle |^{2} }{i\Omega + E_{\nu,{\bf k}+{\bf q}} - E_{\nu',{\bf k}}}\\
								 &\times\Big( f(E_{\nu,{\bf k}+{\bf q}}) - f(E_{\nu',{\bf k}}) \Big),
\end{aligned}
\end{equation}
where $E_{\nu,{\bf k}}$ and $\big|{\bf k},\nu \rangle$ is the $\nu$-th eigenvalue and corresponding eigenvector given by Eq.~\ref{TBHt}.

Fig.~\ref{spinsus} shows the static spin susceptibility $\chi_{s}({\bf q}, 0)$ versus ${\bf q}$ for pure \BFA and \BFP, respectively.
It shows the largest values of static spin susceptibilities in both of these systems occur around ${\bf q} = {\bf Q}\in \{(\pm\pi, 0); (0,\pm\pi)\}$, which is responsible for the scattering vector of the co-linear SDW instability~\cite{Raghu2008, Graser2009}.
At a first glance on Fig.~\ref{spinsus}, we know that the spin-fluctuations in \BFA is much stronger than \BFP from our model parameters.
Since the Stoner condition with onsite Coulomb interaction $U=3.2$ corresponds to  $U\chi_{s}({\bf q}, 0)>1$ for \BFA and $U\chi_{s}({\bf q}, 0)<1$ for \BFP, this implies that the co-linear SDW is stable in \BFA and absent in \BFP.
This feature is consistent with the first principle calculations~\cite{Maier2009, Graser2010} and neutron scattering experiments~\cite{Huang2008, Matan2009, Su2009, Kofu2009, Harriger2011}, as well as Ref.~\onlinecite{ZPYin2014} which also shows the large intensity difference on {\bf Q} for \BFA and \BFP.
If the superconductivity in these compounds is originated from the spin-fluctuations, we can expect that the SC pairing intensity in \BFA should be stronger than that in \BFP, however, the magnetic instability will firstly enter into \BFA.
Therefore, our focus will be to study the phase diagram in the whole doping region ($0<x<1$) of \Px in the next section.
Again, we would like to re-emphasize that our model parameters for \BFA and \BFP are reasonably and qualitatively agreed to the reality from the band structure and static spin-fluctuation studies.

\section{Mean-field calculated phase diagram}
\begin{figure}[htbp]
\subfigure[]{\includegraphics[width=0.48\linewidth]{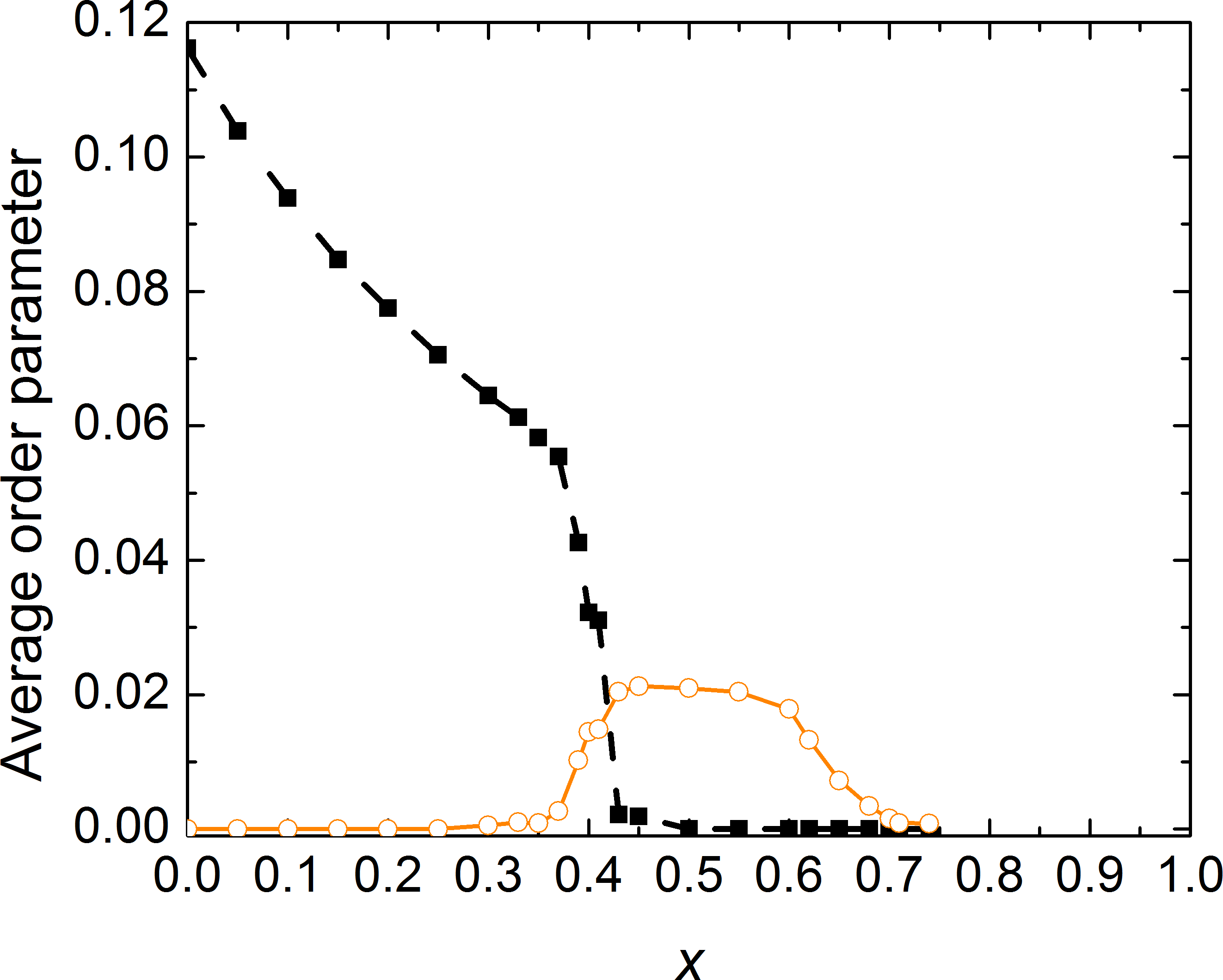}\label{orderpra}}
\;
\subfigure[]{\includegraphics[width=0.48\linewidth]{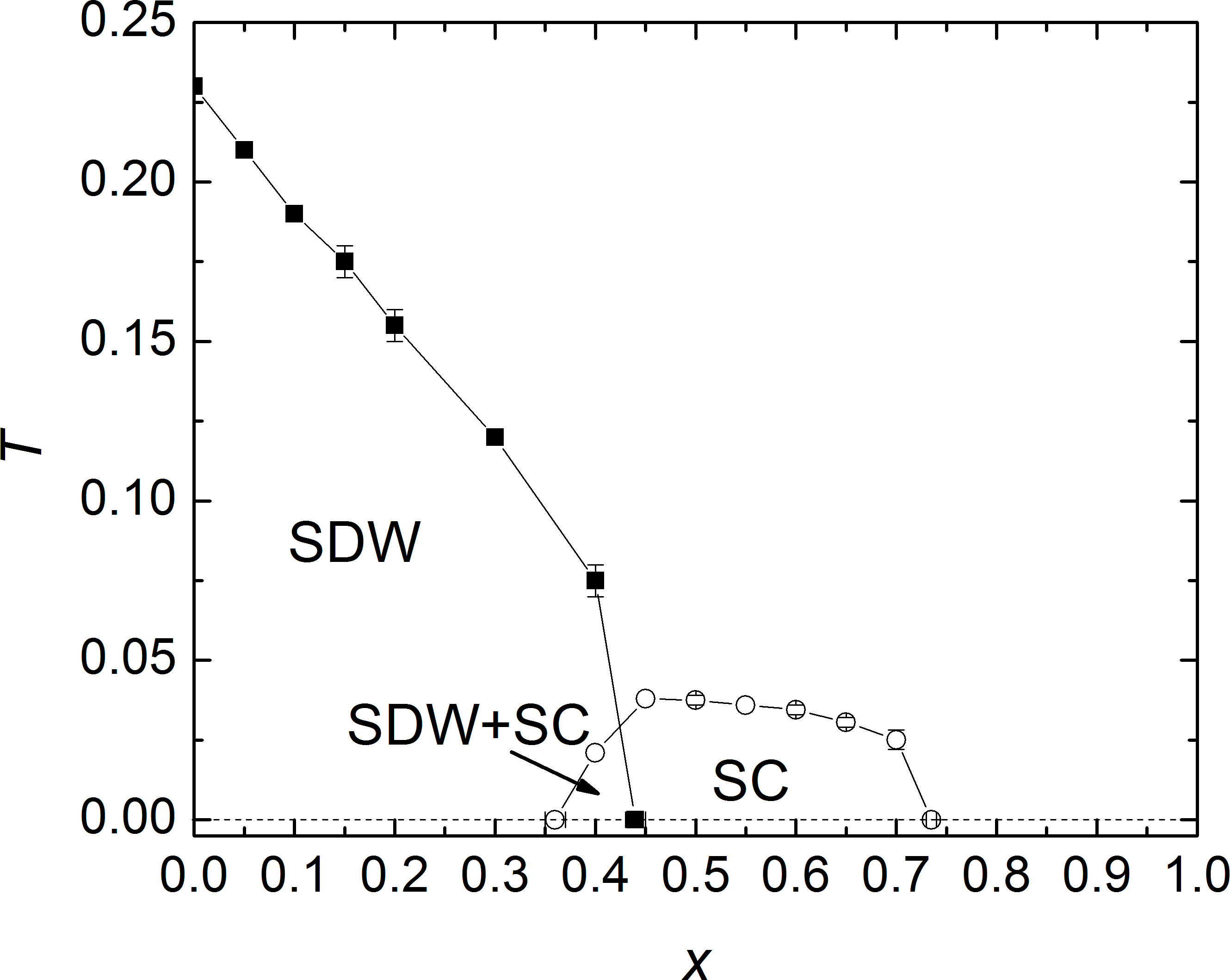}\label{Tphase}}
\caption{(Color online) (a) Phase diagram of averaged order parameters $\langle \big| M \big| \rangle$ and $\langle \big| \Delta_{s} \big| \rangle$ in \Px with respect to different $x$ at $T = 0$, the dash line with black square shows the average SDW order parameter,  while the solid line with circle shows the average SC order parameter, (b) The calculated $T$-$x$ phase diagram of \Px.}\label{phasedia}
\end{figure}

In the following, we study the As-P mixing effect and calculate the phase diagram for \Px.
Based on the mixing parameters described in Eq.~\ref{hoppings} and the Appendix A, we perform the numerical study on a square lattice of $28 \times 28$ sites with periodic boundary conditions, and employ the lattice BdG self-consistent equation given by Eq.~\ref{BdG}. First, we start at $T=0$ for each doping level to obtain the phase boundaries through the site-averaged order parameters, \MAG and \DEL, as shown in Fig.~\ref{orderpra}. Then, we gradually increase $T$ to calculate the temperature dependence on Fig.~\ref{Tphase}. The phase diagrams exhibited in Fig.~\ref{phasedia} are results after making averages over $25$ impurity-configurations. Above or beyond the SDW or SC transition temperature, the relevant averaged order parameters are less than $2\%$ of those magnitudes at $T=0$.  Here the temperature $T$ is in units of the hopping term, $|t_5|$.

\begin{figure}[htbp]
\subfigure[\,x = 0, $\langle \big| M \big| \rangle = 0.11613$]{\includegraphics[width=0.45\linewidth, clip=true]{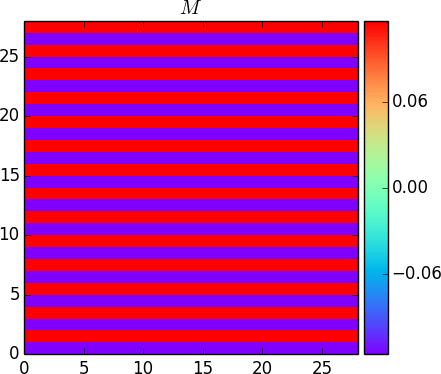}\label{0mag}}
\qquad
\subfigure[\,x = 0.20, $\langle \big| M \big| \rangle = 0.07634$]{\includegraphics[width=0.45\linewidth, clip=true]{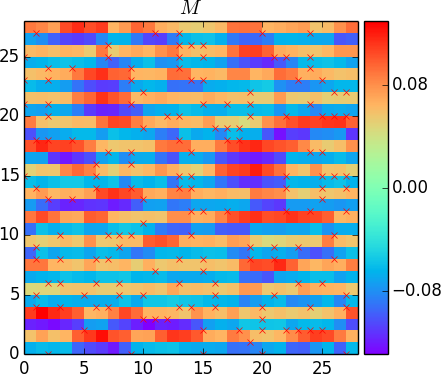}\label{20mag}}
\\
\subfigure[\,x = 0.40, $\langle \big| M \big| \rangle = 0.05059$]{\includegraphics[width=0.45\linewidth, clip=true]{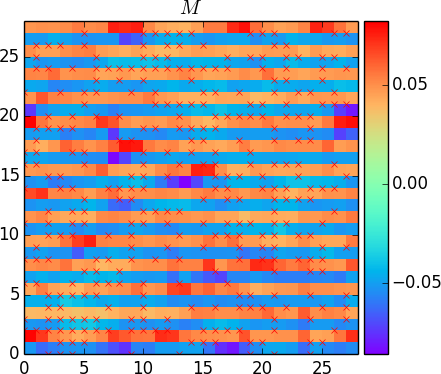}\label{40mag}}
\qquad
\subfigure[\,x = 0.45, $\langle \big| M \big| \rangle = 0.00160$]{\includegraphics[width=0.45\linewidth, clip=true]{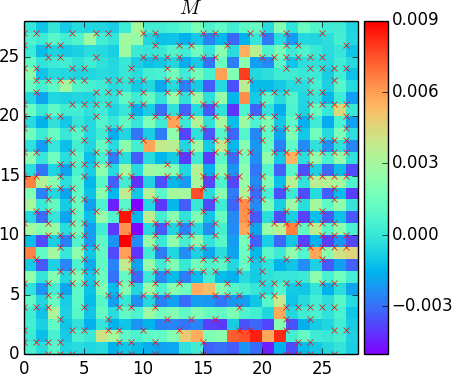}\label{45mag}}
\caption {(Color online) Spatial profiles of local magnetic order parameter $M$ in \Px with $x = 0$, $0.20$, $0.40$ and $0.45$ at $T = 0$, and a specific impurity-configuration is randomly chosen for each $x$. $\times$ (red) represents the position of substituted-P atom.
}
\label{mag}
\end{figure}

\begin{figure}[htbp]
\subfigure[\,x = 0.40, $\langle \Delta_{s} \rangle = 0.01165$]{\includegraphics[width=0.45\linewidth, clip=true]{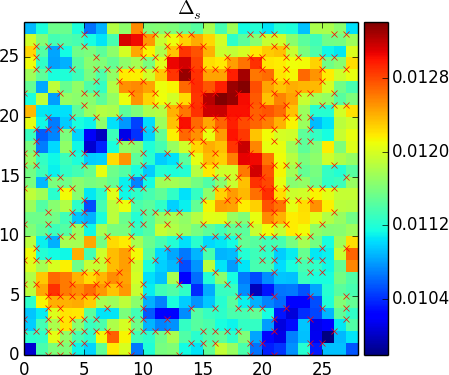}\label{40sup}}
\qquad
\subfigure[\,x = 0.45, $\langle \Delta_{s} \rangle = 0.02125$]{\includegraphics[width=0.45\linewidth, clip=true]{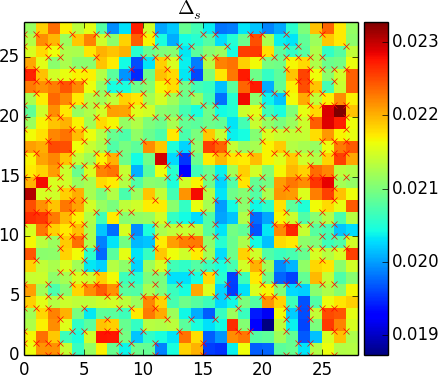}\label{45sup}}
\\
\subfigure[\,x = 0.55, $\langle \Delta_{s} \rangle = 0.02027$]{\includegraphics[width=0.45\linewidth, clip=true]{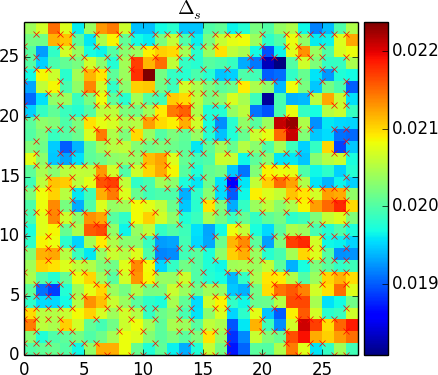}\label{55sup}}
\qquad
\subfigure[\,x = 0.65, $\langle \Delta_{s} \rangle = 0.00798$]{\includegraphics[width=0.45\linewidth, clip=true]{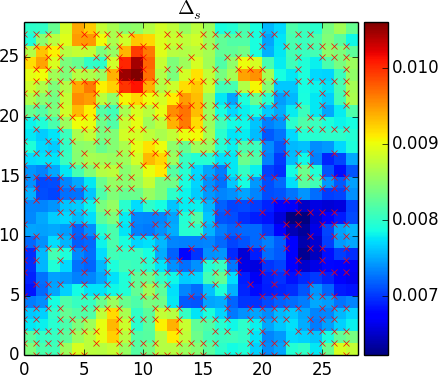}\label{65sup}}
\caption{(Color online) Spatial profiles of local SC order parameter $\Delta_{s}$ in \Px with $x = 0.40$, $0.45$, $0.55$ and $0.65$ at $T = 0$, and a specific impurity-configuration is randomly chosen for each $x$. $\times$ (red) represents the position of substituted-P atom.}\label{super}
\end{figure}

In Fig.~\ref{phasedia}, the pure \BFA is a co-linear SDW state, and the pure \BFP is a paramagnetic metal.
The calculated SC order in doped \Px is much reduced as if we only consider $H_{\tilde t}$ for the kinetic term, and it has been further suppressed if we also take $H_{P_{inter}}$ into account~\cite{footnote02}.
This is not so surprising that the mean-field results of these two pure compounds are expected from our static spin susceptibility study of previous section. In the region, $0<x<0.37$, the SDW order of \BFA is gradually suppressed by the randomly distributed P-substitution. The SC with the $s\pm$-pairing symmetry emerges as a competing order where it has a co-existing region with the SDW order in the region $0.37<x<0.43$ and the SDW order is suddenly dropped to zero at $x~0.43$ as indicated in Fig.~\ref{orderpra}. At $x \simeq 0.45$, the SC order parameter reaches a maximum value, and gradually decreases for $x>0.45$. Finally, the SC order becomes completely suppressed after $x > 0.73$. The overall trend of Fig.~\ref{phasedia} is in good agreement with many experiments~\cite{Jiang2009, Kasahara2010, Nakai2010, Bohmer2012}.

In order to understand the SDW-SC competing effect and together with the As-P mixing picture, we present the spatial images of the local magnetic, $m_i$, and the SC, $\Delta_{i}$, order parameters respectively in Fig.~\ref{mag} and Fig.~\ref{super}. Below each of the graphs of Fig.~\ref{mag} and Fig.~\ref{super} is the averaged value, \MAG and \DEL, respectively. We show several different $x$ at $T = 0$ with a specific configuration of the P-impurities marked by red-$\times$ symbols.

In Fig.~\ref{0mag}, the magnetic order at $x=0$ corresponds to a perfect co-linear SDW phase. These graphs, Fig.~\ref{0mag}-\ref{45mag} and Fig.~\ref{40sup}-\ref{65sup}, demonstrate how the magnetic order weakens as the doping changes for $x=(0,0.2,0.4,0.45)$ and the behavior of the SC dome for $x=(0.4,0.45,0.55,0.65)$. For $x=0.2$, we find that the SDW order is a bit suppressed around the P-impurities as shown in Fig.~\ref{20mag}.
For $x=0.4$, the SDW order is further suppressed as in Fig.~\ref{40mag}, moreover, the SC order now appears as in Fig.~\ref{40sup}.
By comparing Fig.~\ref{40mag} and Fig.~\ref{40sup} ($x=0.4$), we learn that the SDW and SC orders form domains that the stronger region of SC is separated from the stronger region of SDW. For $x=0.45$,  the long range co-linear SDW entirely disappears but there are still short range magnetic order with extremely weak strength as shown in Fig.~\ref{45mag}. On the other hand, Fig.~\ref{45sup} shows the strongest SC on average, we observe that the high intensity spots of the SC order are most likely setting around the As-P boundaries. For $x=0.55$, the SC order is further suppressed and Fig.~\ref{55sup} shows that the high-intensity spots are localized. Finally, when $x$ reaches $0.65$ as shown in Fig.~\ref{65sup}, the SC order is further localized and its high intensity-spots are mostly setting on the P-site free region which tells us that the SC pairing intensity in \BFA is stronger than that in \BFP.

\section{Conclusion}
In conclusion, we have obtained the phase diagrams in the isovalent substituted system \Px with an effective model for the first time. The resulting phase diagrams are in agreement with experiments~\cite{Jiang2009, Kasahara2010, Nakai2010, Bohmer2012}. Here we suppose that the hopping integrals between the Fe sites around the substituted-P are altered. Besides, we assume that there exists a very weak onsite-interorbital scattering on the adjacent four Fe site as well. The suppress of the SDW order in this compound as $x$ increases is caused by the incoherent scattering effect due to the randomly distributed P atoms. The competing SC order starts to show up only when the SDW order becomes significantly weakened. In our calculation, we put the pairing interaction $V_{ij}$ unchanged. However, due to the spin fluctuations, $V_{ij}$ may become smaller as the concentration of substituted-P atoms increases. Unfortunately, it is unclear how to treat this issue in the disordered region. Since $U$ is fixed in our system, as a first order approximation, we can use $t$-$J_{1}$ and -$J_{2}$ model to study the superconductivity in perfect \BFA and \BFP systems. Here $V_{ij} \thicksim J_{2} $ might be weak, but it should not be zero in \BFP. Although we keep the value of $V_{ij}$ around the P-sites unchanged, we also assume a weak onsite inter-orbital scattering-terms near these sites in our calculation. The choice of a smaller $V_{ij}$ around substituted-P would not change the main results of the present paper. Beside, we recognize that there may exist other impurity models for this compound. In fact we have tried several different formalism, and it appears that only the model described in the Appendix A is able to qualitatively account for the experimental phase diagram. With the theoretically obtained phase diagram for \Px system, we should be able to calculate and understand the electronic and thermodynamic properties of this compound at any doping level $x$ (from 0 to 1).

\begin{acknowledgments}
We would like to thank Bo Li, Yao-hua Chen, and Jian Kang for valuable discussions.
This work was supported in part by the Robert A. Welch Foundation under the grant No. E-1146 and the Texas Center for Superconductivity at the University of Houston (Y.Y.Z. \& C.S.T.).
Work at Los Alamos National Laboratory was supported by the U.S. DOE Contract No. DE-AC52-06NA25396 through the LANL LDRD Program (Y.-Y.T.).

\end{acknowledgments}
\section{APPENDICES}
\subsection{\label{suphop} The mixing hopping integrals}

\begin{figure}[htbp]
\includegraphics[width=0.5\linewidth]{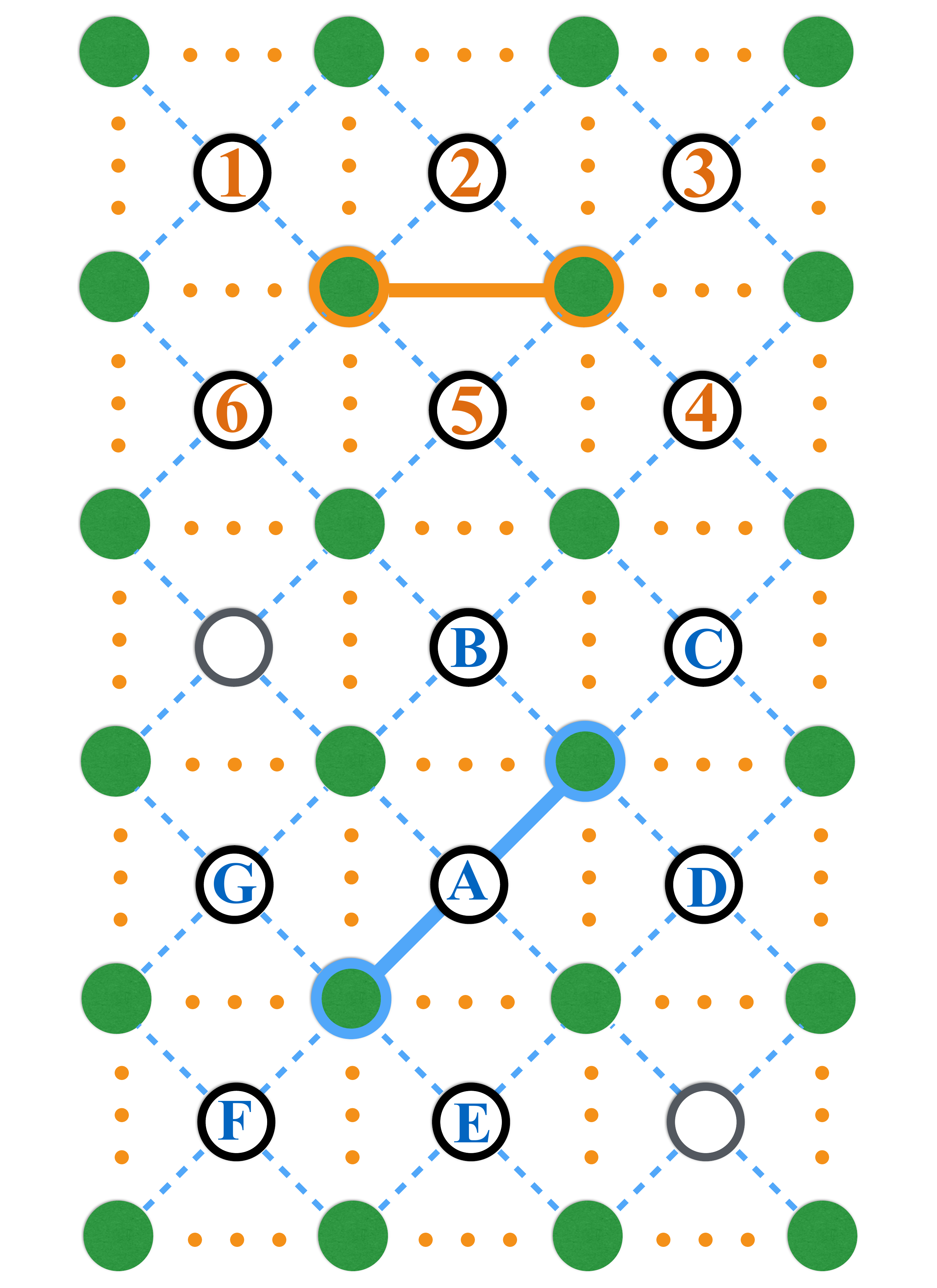}
\caption{Schematic picture for the concept of mixing NN and NNN hopping terms for \Px.} \label{mix}
\end{figure}

Here, we give the details about how the hopping integrals, $\tilde t$, mixing with the existence of the substituted-$P$ atoms in the \Px system.
The six hopping integrals are $t_{1-6}$ for pure \BFA (see Refs.~\onlinecite{YuanYenepl} and~\onlinecite{HuaChenprb}) and $t'_{1-6}$ for pure \BFP.
The details of $t_{1-6}$ and $t'_{1-6}$ are given in the present manuscript.
We group the mixing hopping integrals into two categories: {\bf i)} NN hopping integrals, {\bf ii)} NNN hopping integrals, as shown in Fig.~\ref{mix}.
In Fig.~\ref{mix}, the green-solid circle present for the Fe atoms, the open circle present for either As or substituted P atoms, the orange-dotted line is the NN hopping terms ($\tilde t_{1,5}$), the red-dashed line is the NNN hopping terms ($\tilde t_{2,3,4}$); we use numbers (1 ... 6) to locate the atoms which affect the nearby NN hopping term (orange-solid line) and alphabets (A ... G) to locate the atoms which affect the nearby NNN hopping term (red-solid line).

Now we consider all the situations for the mixing hopping terms for case {\bf i)} and {\bf ii)},\\
{\bf i) NN hopping integrals} (the orange-solid line in Fig.~\ref{mix}):
\begin{equation}\label{NN}
\tilde t_{1,\,5} = \left\{ \begin{split}
                            &\mbox{1) Both As atoms in 2 and 5 are substituted by P:} \\
                            &\qquad t^{\prime}_{1,\,5}, \\
                            &\mbox{2) Only one As atom in 2 or 5 is substituted by P:} \\
                            &\qquad a\, t_{1,\,5} + b\, t^{\prime}_{1,\,5},\\
                            &\mbox{3) At least one As in 1, 3, 4 or 6 is substituted by P:} \\
                            &\qquad a^{\prime} t_{1,\,5} + b^{\prime} t^{\prime}_{1,\,5},\\
                            &\mbox{4) All of them (1 to 6) are As atoms:} \\
                            &\qquad t_{1,\,5}.
                        \end{split}
                   \right.
\end{equation}
where $a = 1/4$, $b = 3/4$ and $a^{\prime} = 1/2$, $b^{\prime} = 1/2$.\\
{\bf ii) NNN hopping integrals} (the red-solid line in Fig.~\ref{mix}):
\begin{equation}\label{NN}
\tilde t_{2,\,3,\,4} = \left\{ \begin{split}
                            &\mbox{1) the As atoms in A are substituted by P:} \\
                            &\qquad t^{\prime}_{2,\,3,\,4}, \\
                            &\mbox{2) At least one As atom in B-G is substituted by P:} \\
                            &\qquad a\, t_{2,\,3,\,4} + b\, t^{\prime}_{2,\,3,\,4},\\
                            &\mbox{3) All of them (A to G) are As atoms:} \\
                            &\qquad t_{2,\,3,\,4}.
                        \end{split}
                   \right.
\end{equation}
where $a = 1/4$, $b = 3/4$.
Note that there are two directions of the NN (NNN) hopping integrals which corresponds to two different configurations of the labeling, 1 to 6 (A to G); here we only show one of it.

In our previous discussions, we suppose there exist different hopping integrals between all and some certain substituted-P atoms. The ratios of $a$ ($b$) and $a^{\prime}$ ($b^{\prime}$) are chosen arbitrarily, originally, if we choose $a = 1/2$, and $a^{\prime} = 1$, the SDW will emerge at a big value of $x \approx 0.5$. We have tried different sets of ratios, and these values, chosen in the paper, could give qualitatively comparable experimental phase diagram.

\subsection{\label{supfolded} Band structure and Fermi surface of the 2-Fe per unit cell Brillouin Zone}
\begin{figure}[htbp]
\subfigure[]{\includegraphics[width=0.4\linewidth, clip=true]{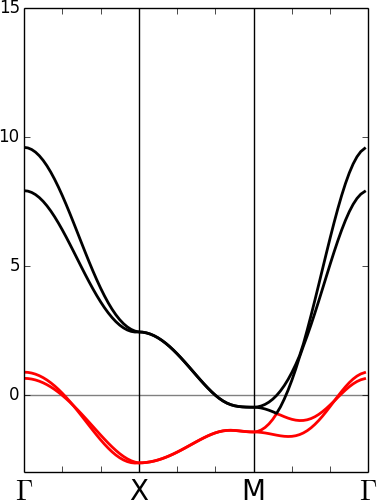}}
\qquad\quad
\subfigure[]{\includegraphics[width=0.4\linewidth, clip=true]{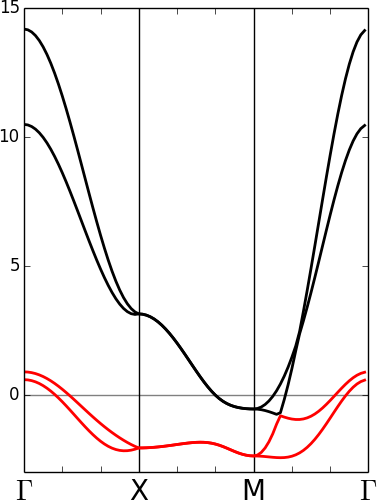}}
\\
\!\!\!\!\!\!\!\!
\subfigure[]{\includegraphics[width=0.42\linewidth, clip=true]{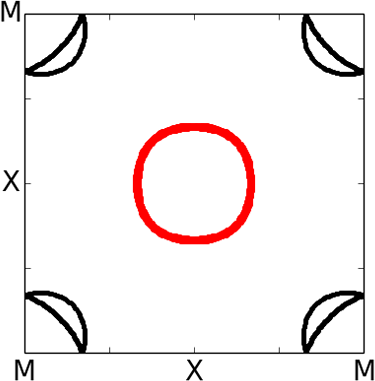}}
\qquad
\subfigure[]{\includegraphics[width=0.42\linewidth, clip=true]{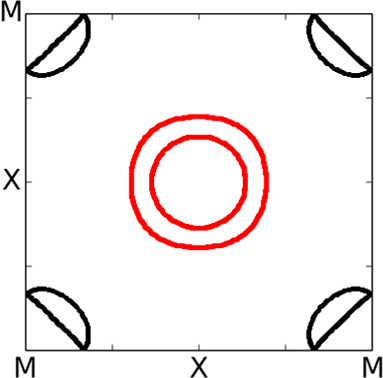}}
\caption{\label{folded}
(Color online) The model calculated band structure of two-orbital model on the 2-Fe per unit cell Brillouin Zone (BZ) for (a) \BFA ($x=0$), (b) \BFP ($x=1$); and Fermi surface on the 2-Fe per unit cell BZ for (c) \BFA ($x=0$), (d) \BFP ($x=1$). Here the fermi energy is shifted to zero for $1/2$ filling.}
\end{figure}

In Fig.~\ref{folded}, we show the band structures and Fermi surfaces of the pure \BFA and \BFP systems in the BZ of 2-Fe atoms per unit cell respectively.


\begin{references}
\bibitem{footnote03}
In Ref.~\onlinecite{BLi}, mathematically, we can also find another mean-field decoupled scheme for the interaction $H_\Delta$,  which is $\sum_{ij}\langle n_i \rangle n_j$.
This mean-field decoupled order is usually responsible for a charge-density-wave (CDW), however, we did not find that such an order is stable.
\end{references}
\end{document}